\documentclass[%
aip,
amsmath,amssymb,
reprint,%
]{revtex4-2}

\usepackage{graphicx}
\usepackage{dcolumn}
\usepackage{bm}

\usepackage{mathtools}

\usepackage[utf8]{inputenc}
\usepackage[T1]{fontenc}
\usepackage{mathptmx}
\usepackage{etoolbox}
\usepackage{subfigure}
\usepackage{hyperref}
\hypersetup{hypertex=true,
	colorlinks=true,
	linkcolor=cyan,
	anchorcolor=cyan,
	citecolor=cyan,
	urlcolor=cyan}
\usepackage{color}
\makeatletter
\def\@email#1#2{%
	\endgroup
	\patchcmd{\titleblock@produce}
	{\frontmatter@RRAPformat}
	{\frontmatter@RRAPformat{\produce@RRAP{*#1\href{mailto:#2}{#2}}}\frontmatter@RRAPformat}
	{}{}
}%
\makeatother
\begin{document}
	\title{Voltage and power-frequency electric field measurements with Rydberg-atom interferometry}

\author{Yingying Han}
\email[Corresponding email: ]{hanyingying@sztu.edu.cn}
\affiliation{Shenzhen Key Laboratory of  Ultraintense Laser and Advanced Material Technology, Center for Intense Laser Application Technology, and College of Engineering Physics, Shenzhen Technology University, Shenzhen 518118, China}

\author{Changfa He}
\affiliation{Shenzhen Key Laboratory of  Ultraintense Laser and Advanced Material Technology, Center for Intense Laser Application Technology, and College of Engineering Physics, Shenzhen Technology University, Shenzhen 518118, China}

\author{Zhenxiong Weng}
\affiliation{Shenzhen Key Laboratory of  Ultraintense Laser and Advanced Material Technology, Center for Intense Laser Application Technology, and College of Engineering Physics, Shenzhen Technology University, Shenzhen 518118, China}

\author{Peng Xu}
\affiliation{School of Physics, Zhengzhou University, Zhengzhou 450001, China}
\affiliation{Institute of Quantum Materials and Physics, Henan Academy of Sciences, Zhengzhou 450046, China}

\author{Yanting Zhao}
\affiliation{State Key Laboratory of Quantum Optics Technologies and Devices, Institute of Laser Spectroscopy,
Shanxi University, Taiyuan 030006, People’s Republic of China}
\affiliation{Collaborative Innovation Center of Extreme Optics, Shanxi University, Taiyuan 030006, People’s Republic of China}

\author{Tao Wang}
\email[Corresponding email: ]{tauwaang@cqu.edu.cn}
\affiliation{Department of Physics, and Center of Quantum Materials and Devices, Chongqing University, Chongqing 401331, China}
\affiliation{Center of Modern Physics, Institute for Smart City of Chongqing University in Liyang, Liyang 213300, China}	
	\date{\today}%
\begin{abstract}
We present a Rydberg-atom interferometry-based technique for voltage measurement between electrodes embedded in an atomic vapor cell, enabling the detection of weak voltages ($<0.1$V) and unambiguous discrimination between positive and negative polarities. This makes up for the shortcomings of measurements based on the Stark effect, which suffer from quadratic field dependence (limiting sensitivity in weak-field regimes) and incapable of distinguishing the electric field direction. Furthermore, this method extends naturally to power-frequency (PF) electric field measurements by exploiting the quasi-static approximation – valid given the PF field's characteristic timescale ($\sim10^{-2}$s) vastly exceeds the interferometric measurement duration ($\sim10^{-6}$s). Crucially, our protocol provides instantaneous PF field reconstruction, providing comprehensive information including amplitude, frequency and phase. These advancements have direct implications for traceable voltage measurements and non-invasive characterization of PF fields near high-voltage infrastructure.	
\end{abstract}
\maketitle
Rydberg atoms are highly sensitive to external electric fields due to their large polarizability, making them an excellent platform for electric field measurements\cite{WOS000803107000014,WOS001065914200001,WOS000310836700021,WOS000537039500003}. The basic principle of electric field measurements relies on the Stark effect, where the energy levels of Rydberg states shift or split in the presence of an electric field. This shift can be precisely measured using techniques like electromagnetically induced transparency (EIT) spectroscopy \cite{PhysRevLett.98.113003,PhysRevA.100.063427,WOS:000375846600053,WOS:001245518300001}. Rydberg atoms offer significant advantages for electric field measurements, including exceptional sensitivity, a broad frequency response spanning from (direct current) DC to THz, and the unique capability of self-calibration, making them a powerful tool for precision metrology in diverse electromagnetic environments~\cite{PhysRevApplied.18.014033,PhysRevApplied.13.054034,PhysRevApplied.18.014045,dingNP,WOS:001089501700010}.

The DC electric field strength $\epsilon_{dc}=V/d$ between two parallel plates (separation $d$) in an atomic vapor cell is determined by measuring the DC Stark shifts in Rydberg spectra, which subsequently allows calibration of the induced voltage \cite{PhysRevLett.82.1831,2022Electromagnetically,2009Enhanced,PhysRevApplied.18.024001,PhysRevResearch.6.023138,Ma:20}. However, due to the quadratic dependence of the Stark shift on electric field strength, these methods inherently cannot determine the DC field direction and exhibit diminished sensitivity in weak field regimes. The DC field direction affects device performance and measurement accuracy in high-precision applications, such as influencing thermal effects in crystal materials~\cite{PhysRevMaterials.8.094408} and the trajectories and energy distribution of electrons in plasma physics~\cite{Gedalin_2024}. Additionally, weak DC fields require characterizations and they are measured indirectly by fitting the Rydberg spectra~\cite{PhysRevResearch.6.023138,2009Enhanced}, which inherently rely on model-dependent assumptions. Consequently, there is a pressing need for another direct scheme capable of measuring weak DC fields and distinguishing the direction of the DC fields.

In this work, we investigate the use of Rydberg-atom interferometry to detect weak DC fields (voltages) and distinguish the directions of the DC fields (positive and negative polarities). The interference effect caused by Floquet EIT in radio-frequency (RF) dressed Rydberg atoms is the core mechanism of this scheme~\cite{han2025}. By scanning the RF phase, the sideband of the Rydberg spectra exhibits a bimodal structure, and the peak separation of the bimodal structure is monotonically decreasing as the voltage increases from $-0.1$V/cm to $0.1$V/cm. It is noteworthy that the peak separation varies linearly with the voltage near zero strength, which will greatly improve the measurement ability of the voltage. Given the ability to distinguish the DC field direction, we can reasonably discretize the slowly varying power-frequency (PF) field into a large number of quasi-DC fields. This method allows us to characterize the dynamics of the PF field and extract its frequency, amplitude, and phase.

\begin{figure}[thp]
\setlength{\abovecaptionskip}{0.cm}
\setlength{\belowcaptionskip}{-0.cm}
\includegraphics[width=3in]{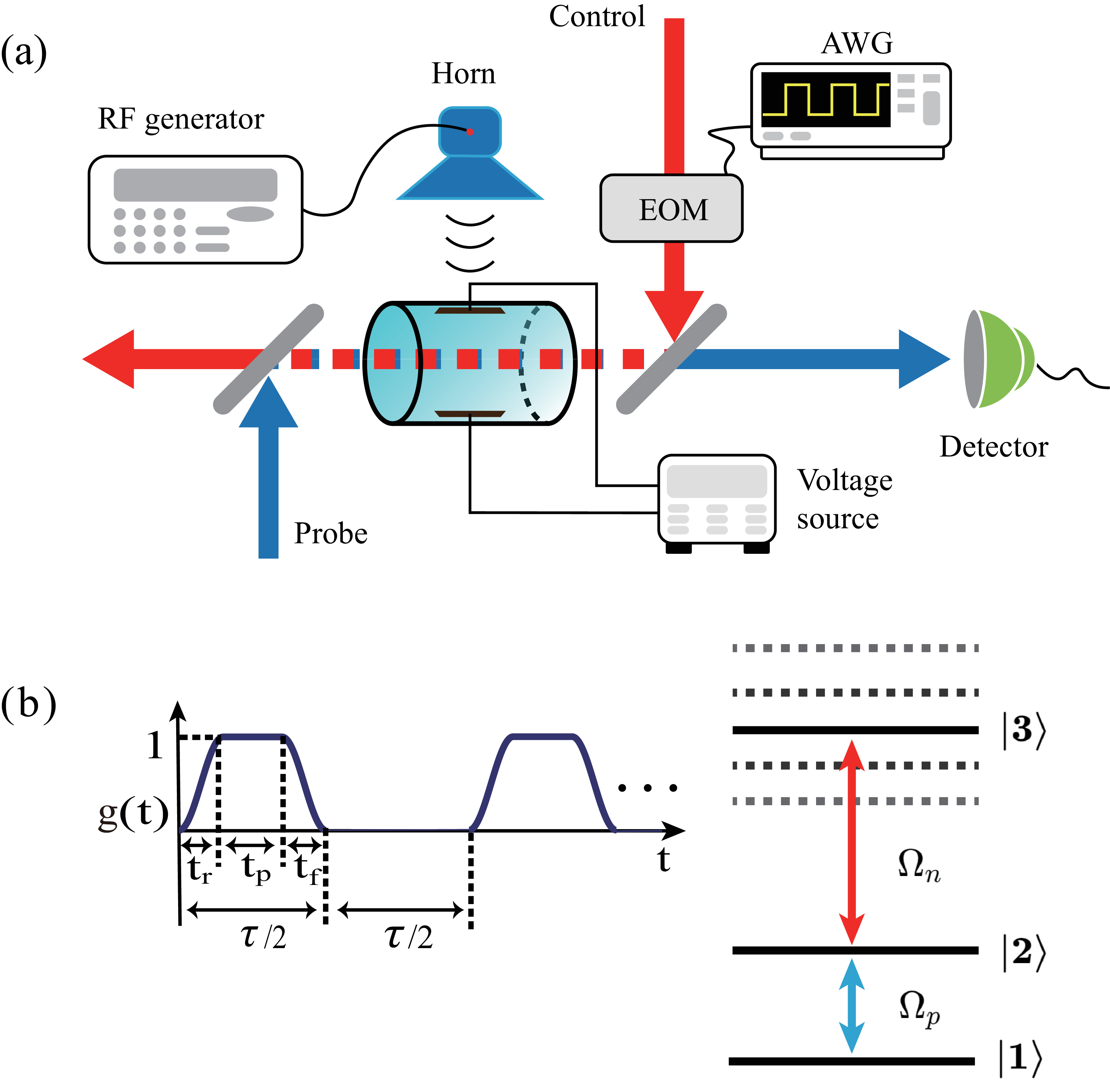}
\caption{\label{fig:1}(a) Schematic diagram of the experimental setup. (b) A three-level system is driven by a resonant probe field with strength $\Omega_p$ and a detuned control field with strength $g(t)\Omega_c$. The periodic function $g(t)$ with period $\tau=2(t_r+t_p+t_f)$ consists of a square wave with a duty cycle of 0.5, rise, fall and maximum-amplitude (i.e., 1) duration times $t_r$, $t_f$ and $t_p$, respectively. The Rydberg state splits into a series of sideband under the modulations of both periodic function $g(t)$ and RF field. The coupling strength between the $n$-th sideband and state $|2\rangle$ is $\Omega_n$.}
\end{figure}

Schematic of the vapor cell setup with the probe and control beams counter-propagating through the vapor
cell, and the control beam passes through an electro-optic modulator (EOM) modulated by an arbitrary waveform generator (AWG), as shown in Fig.~\ref{fig:1}(a). Inside the cell are two stainless-steel parallel electrodes with separation $d$. The to-be-measured voltage is applied to the two electrode plates. A MHz RF field is applied using a horn antennas fed by a RF generator. The transmission of the probe beam through the cell is monitored by the detector. The energy-level scheme is a three-level system with ground state $|1\rangle$, excited state $|2\rangle$ and Rydberg state $|3\rangle$, as shown in Fig.~\ref{fig:1}(b). A weak probe field couples transition $|1\rangle\leftrightarrow|2\rangle$ with coupling strength $\Omega_p$, and a strong control field couples transition $|2\rangle\leftrightarrow|3\rangle$ with periodically modulated coupling strength $g(t)\Omega_c$. $g(t)$ is generated by AWG and it is a periodic function with frequency $\omega_g$ and initial phase $\Phi_g$. The Rydberg state is highly sensitive to electric fields. Its energy is shifted by both the DC field (applied by the voltage between electrodes) and the RF field, whose frequency (in the tens of MHz range) is far from neighboring Rydberg transitions. Then the energy of the Rydberg state is  $\omega_3(t)=\omega_3-\alpha\epsilon^2(t)/(2\hbar)$ with $\alpha$ the dipole polarizability of the Rydberg state and $\epsilon(t)=\epsilon_{dc}+\epsilon_{rf}\text{cos}(\omega_st+\Phi)$ with DC intensity $\epsilon_{dc}$, RF amplitude $\epsilon_{rf}$, phase $\Phi$ and frequency $\omega_s$.  The Hamiltonian of the system can be expressed as ($\hbar=1$ in this work) 
    \begin{align}\label{eq:hami}
	\hat{H}(t)&=\omega_{1}|1\rangle\langle1|+\omega_{2}|2\rangle\langle2|+\omega_{3}(t)|3\rangle\langle3|\notag\\
&-\frac{1}{2}\left[ \Omega_{p}\left( e^{i\omega_{p}t}+e^{-i\omega_{p}t}\right) |1\rangle\langle2|+\text{H.c.}\right] \notag\\
	       &-\frac{1}{2}\left[ g(t)\Omega_c\left( e^{i\omega_{c}t}+e^{-i\omega_{c}t}\right)|2\rangle\langle3|+\text{H.c.}\right],
    \end{align}
where $\omega_{p}$ and $\omega_{c}$ are the frequencies of the probe and control fields, respectively. In this scheme, either the RF field or the periodically modulated control field can split the Rydberg state into a series of sidebands. When both cases are present, there is generally no longer a stable sideband, which makes the situation complicated. However, if there is an integer multiple relationship between the frequencies of the RF field and the periodic function $g(t)$, there will be a series of stable sidebands. Here we consider the situation that: $\omega_s=L\omega_g$ with integer $L$. 

We transform the system into the interaction picture and consider the conditions which makes the sidebands resolvable~\cite{liu:cpb,2009Enhanced}, as shown in Fig.~\ref{fig:1}(b). When the control field dutuning is close to one of the sidebands, we can neglect the other non-resonant terms and make the rotating wave approximation (RWA)~\cite{QO}. Then the Hamiltonian just for the $n$-th band (marked by $|n\rangle$) can be simplified to (see supplementary material)  
\begin{align}\label{eq:rwa}
	\hat{H}^{n}_{\text{RWA}}(t)&=
	-\Delta_p|2\rangle\langle2|-(\Delta_p+\Delta_n)|n\rangle\langle n| \notag\\
  &+\Omega_p/2|1\rangle\langle2|+\Omega_n/2|2\rangle\langle n|+\text{H.c.}.
\end{align}	
where
\begin{equation}\label{eq:omen}
	\Omega_{n}=\Omega_{c}\sum_{m=-\infty}^{\infty}g_{n-mL}A_{m}e^{im(\Phi-L\Phi_g)},
\end{equation}
with $ A_{m}=\sum_{k=-\infty}^{+\infty}J_k\left(\frac{\alpha\varepsilon_{rf}^{2}}{8\omega_{s}}\right)
	J_{m-2k}\left(\frac{\alpha\varepsilon_{dc}\varepsilon_{rf}}{\omega_{s}}\right)$,
$\Delta_{p}=\omega_{p}-\omega_{21}$, $\Delta_{n}=\Delta_{c}+\omega_\alpha+n\omega_g$ and $\Delta_{c}=\omega_{c}-\omega_{32}$. $J_k(.)$ is the first kind Bessel function. $\omega_{\alpha}=\alpha\epsilon_{dc}^{2}/2+\alpha\epsilon^{2}_{rf}/4$ is the Stark shift due to the DC and RF fields. $g_{n}$ is the Fourier coefficient of the periodic function $g(t)$. In this work, square-wave modulation is considered and the Fourier coefficients of it are shown in supplementary material. From Eq.~(\ref{eq:omen}), we see that $\Omega_n$ is the sum of a series of terms with different phases. The transmission of the probe field is proportional to $|\Omega_n|^2$, where the cross terms represent the interference among multi-paths with different phases. We refer to this measurement scheme, which utilizes interference effects in the Rydberg atom system, as Rydberg-atom interferometry.

\begin{figure}[thp]
\setlength{\abovecaptionskip}{0.cm}
\setlength{\belowcaptionskip}{-0.cm}
\includegraphics[width=3.4in]{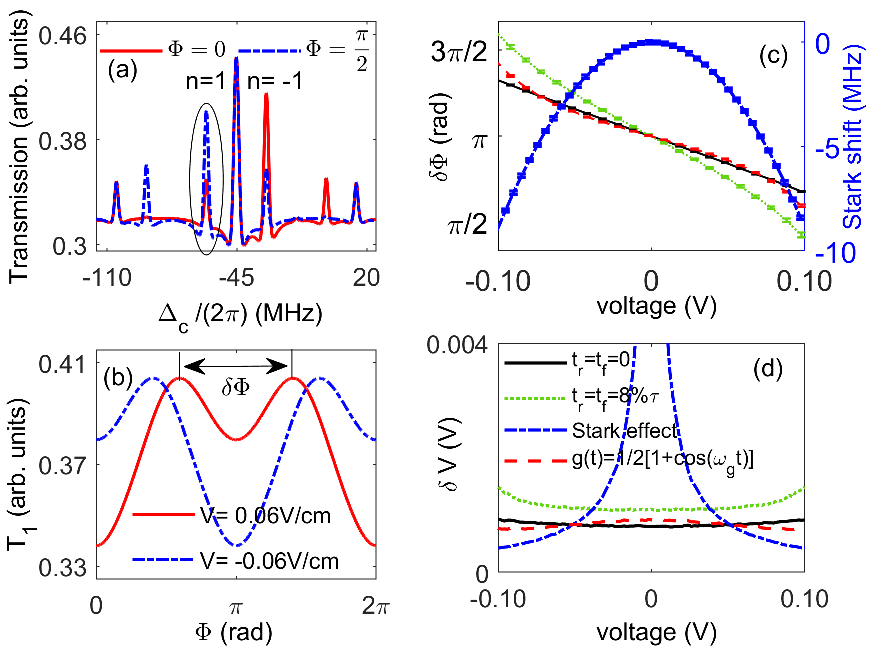}
\caption{\label{fig:2}(a) The transmission of the probe field against the control field detuning $\Delta_c$ under different RF phase with $V=0.02$V. (b) The maximum transmission of $n=1$ band (marked by $\text{T}_1$) varies with RF phase under $V=\pm0.06$V. (c) The $\delta \Phi$ varies with voltage under different periodic functions. For comparison, we also display the Stark shift caused by the DC field ($\epsilon_{dc}=V/d$ with $d=2$mm), as the blue dash-dotted line shown. Here all the curves with error bars are created by repeating the process 10,000 times while taking into account 100KHz fluctuations of $\omega_p$ and $\omega_c$. The legends are the same as in (d). Note that for the square-wave modulation (i.e., black solid line with $t_r=t_f=0$ and green dotted line with $t_r=t_f=8\%\tau$) $\omega_g/(2\pi)=15~\text{MHz}$ and the cosine modulation (red dashed line) $\omega_g/(2\pi)=30~\text{MHz}$. (d) The accuracy $\delta V$ as a function of voltage. } 
\end{figure}

To study the transmission of the probe filed, we investigate the susceptibility $\chi$, the imaginary part of which indicates the absorption characteristic of the medium. Considering atomic thermal motion, the Doppler effect shifts the probe field frequency to $\omega_{p}\left(1+\nu/c\right)$ and the control field frequency to $\omega_{c}\left(1-\nu/c\right)$ for atoms moving at velocity $\nu$, where $c$ is the speed of light. Then the Doppler averaged susceptibility for the probe field~\cite{berman2011principles,PhysRevA.51.576}
\begin{align}\label{eq14}
	\chi=\int \frac{iN_{0}|d_{12}|^2e^{-v^2/u^2}/(\sqrt{\pi}u\epsilon_{0})}{\Gamma_{1}-i\Delta_{p}-i\omega_{p}\frac{\nu}{c}+\frac{|\Omega_{n}|^{2}/4}{\Gamma_{2}
-i(\Delta_{p}+\Delta_{n})-i(\omega_{p}-\omega_{c})\frac{\nu}{c}}}d\nu,
\end{align}
where $u=\sqrt{2k_{b}T_{0}/m}$ is the most probable velocity with atom mass $m$. $k_{b}$ is the Boltzmann constant. $T_{0}$ is medium temperature. $d_{12}$ is the dipole matrix element corresponding to the transition $|1\rangle\to |2\rangle$. $N_{0}$ and $\epsilon_{0}$ are the atom number density and the vacuum permittivity, respectively. Then we can obtain the $n$-th band transmission $T_{n}$ of the probe laser as it passes through the medium	
\begin{align}\label{eq16}
	T_{n}=\mathrm{exp}\left[-2\pi lIm(\chi)/\lambda_{p}\right],
\end{align}	
where $l$ represents the distance the probe field travels through the medium, and $\lambda_{p}$ is the wavelength of the probe field. Finally, we can get the whole Rydberg spectra by summing the contributions from all the sidebands $T^{\text{Tot}}=\sum_{n}T_{n}$. From Eqs.~(\ref{eq14})-(\ref{eq16}), we see that the transmission of the probe field is determined by $|\Omega_n|^2$, which is a function of $\epsilon_{dc}$ and RF phase $\Phi$. Due to the precise phase modulation capability of the RF field~\cite{PhysRevLett.127.033601,NP21,WOS:001089501700010}, we can investigate the relationships between the Rydberg spectra and the DC fields by scanning the RF field phase. This approach enables us to propose a DC field measurement scheme fundamentally distinct from the Stark effect.

We present a specific example for the measurement of voltage induced by two parallel electrodes with separation $d$ based on the Rydberg-atom interferometry. According to Ref.~\cite{2022Electromagnetically}, we set $d=2$ mm in this work. Considering a $\prescript{133}{}{\text{Cs}}$ three-level system: $6S_{1/2}-6P_{3/2}-53S_{1/2}$ in a vapor cell at a controlled temperature of $300$K. The polarizability of the Rydberg state $53S_{1/2}$ is $\alpha=71.1\hbar~\text{MHz}/(\text{V/cm})^2$\cite{high}. The decay parameters of the system are $\Gamma_1/(2\pi)=2.65~\text{MHz}$ and $\Gamma_2/(2\pi)=0.015~\text{MHz}$. The wavelengths
of probe and control laser are chosen at $\lambda_p = 852$ nm and $\lambda_c= 510$ nm to induce the corresponding transitions, respectively. In this work, the RF field as a reference field with frequency $\omega_s/(2\pi)=30~\text{MHz}$ and amplitude $\epsilon_{rf}=4$ V/cm, $\Omega_p/(2\pi)=0.1$MHz, $\Omega_c/(2\pi)=5$MHz, and we set the square-wave frequency to the half that of the RF field, i.e., $L=2$. 

In Fig.~\ref{fig:2}(a), we show the transmission of the probe field varies with the control field detuning under different RF phase. We see that the interval between the adjacent bands is $\omega_{g}/(2\pi)=15$ MHz, and the main peak is shifted by the Stark shift $\omega_{\alpha}/(2\pi)\approx-45\text{MHz}$. The maximum transmissions of the first-order bands (marked by $\text{T}^{\text{max}}_{n=\pm1}$, $\text{T}_{\pm1}$ for short) varies dramatically with the RF phase. By fixing the control field detuning at the peak of the $n=1$ band and scanning the RF phase, the variation of $\text{T}_1$ with RF phase is obtained, as shown in Fig.~\ref{fig:2}(b). For $V=\pm0.06$V, both curves show bimodal structure, but the peak separations (marked by $\delta\Phi$) are different. This indicates that the positive and negative polarity can be distinguished. Note that the peak separations $\delta\Phi$ are the same for different voltage under the square-wave modulation with $L=1$ (more details see supplementary material).

To further verify the above results and discuss the systematic effects, we consider the fluctuations of $\omega_p$ and $\omega_c$. Here we ignore the fluctuation of the coupling strength of the probe and control fields, i.e., $\Omega_p$ and $\Omega_c$, because $\delta \Phi$ is independent of $\Omega_p$ and $\Omega_c$ (see supplementary material). In Fig.~\ref{fig:2}(c) we show the peak separations $\delta{\Phi}$ changing with voltage under different periodic functions. The results are calculated by repeating the process to obtain $\delta\Phi$ 10,000 times while taking into account 100KHz fluctuations of $\omega_p$ and $\omega_c$, and the error bars in Fig.~\ref{fig:2}(c) are the standard deviations of $\delta\Phi$. The black solid line and the green dotted line represent the square-wave modulation with $t_r=t_f=0$ and $t_r=t_f=8\%\tau$, respectively, and $\omega_g/(2\pi)=15$ MHz. The red dashed line is a commonly used cosine modulation in experiments, with optimal modulation frequency $\omega_g/(2\pi)=30$ MHz. These curves demonstrate that $\delta \Phi$ monotonically decreases as the voltage increases from $-0.1$V to $0.1$V, confirming the ability to distinguish between positive and negative polarities using interference-based measurements. Notably, the peak separation $\delta\Phi$ varies linearly with the voltage near zero strength. In contrast, measurements relying on the Stark effect cannot achieve this distinction and fail near weak voltages close to 0, as illustrated by the blue dash-dotted line in Fig.~\ref{fig:2}(c). For the Stark shift, the control field is not modulated and the RF field is useless. Thus the results for the Stark shift are calculated by replacing $\Omega_n$ and $\Delta_n$ in Eq.~(\ref{eq14}) with $\Omega_c$ and $\Delta_c'=\omega_c-\omega_{32}+\alpha\epsilon_{dc}^2/2$. We further demonstrate the accuracy of the voltage measurement in Fig.~\ref{fig:2}(d). Here the accuracy defined as $\delta V=\delta(\delta \Phi)/[\partial(\delta \Phi)/\partial V]$ with standard deviation $\delta(\delta \Phi)$. The standard deviations $\delta(\delta \Phi)$ are the same as the error bars in Fig.~\ref{fig:2}(c). It is revealed that the minimum resolvable voltage near zero using the Rydberg-atom interferometry scheme is significantly lower than that achievable with the Stark-shift-based scheme. These findings highlight the superior sensitivity of Rydberg-atom interferometry for precision voltage measurements, particularly in the vicinity of zero voltage.

\begin{figure}[thp]
\setlength{\abovecaptionskip}{0.cm}
\setlength{\belowcaptionskip}{-0.cm}
\includegraphics[width=3.4in]{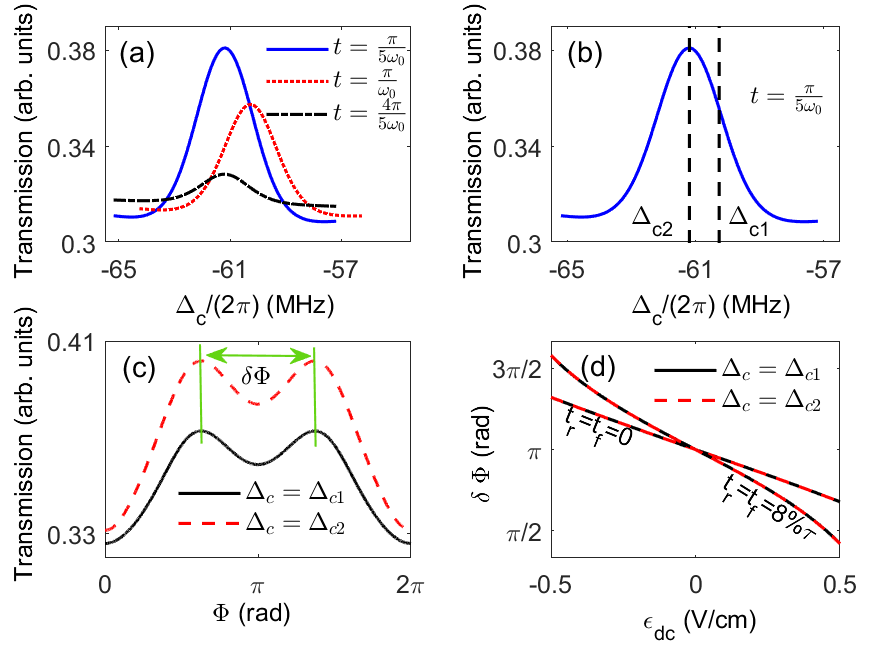}
\caption{\label{fig:3}(a) The $n=1$ band in Fig.~\ref{fig:2}(a) change with the time-varying PF field at different times with PF frequency $\omega_0$: $t=\pi/(5\omega_0)$ for the blue solid line, $t=\pi/(\omega_0)$ for the black dash-dot line, and $t=4\pi/(5\omega_0)$ for the red dotted line. The RF phase is $\Phi=\pi$. (b) The $n=1$ band with $\epsilon_{dc}=\epsilon_{pf}(t)|_{t=\pi/(5\omega_0)}$. The black dashed lines are $\Delta_{c1}=-\alpha\epsilon_{rf}^2/4-\omega_g$ and $\Delta_{c2}=-\alpha\epsilon_{rf}^2/4-\alpha\epsilon_{dc}^2/2-\omega_g$ with $\epsilon_{dc}=\epsilon_{pf}(t)|_{t=\pi/(5\omega_0)}$, respectively. The RF phase is $\Phi=\pi$. (c) The transmissions of the probe field vary with the RF phase under fixed control field detuning $\Delta_{c1}$ and $\Delta_{c2}$, respectively. (d) $\delta \Phi$ vary with $\epsilon_{dc}$ under fixed control field detuning $\Delta_{c1}$ and $\Delta_{c2}$, respectively. In this figure, PF phase $\Phi_{pf}=0$, PF amplitude $\epsilon_0=0.5$V/cm and the periodic function $g(t)$ is square wave with frequency $\omega_{g}/(2\pi)=15\text{MHz}$ and $t_r=t_f=0$ in (a)-(c).}
\end{figure}

\begin{figure}[thp]
\setlength{\abovecaptionskip}{0.cm}
\setlength{\belowcaptionskip}{-0.cm}
\includegraphics[width=3.4in]{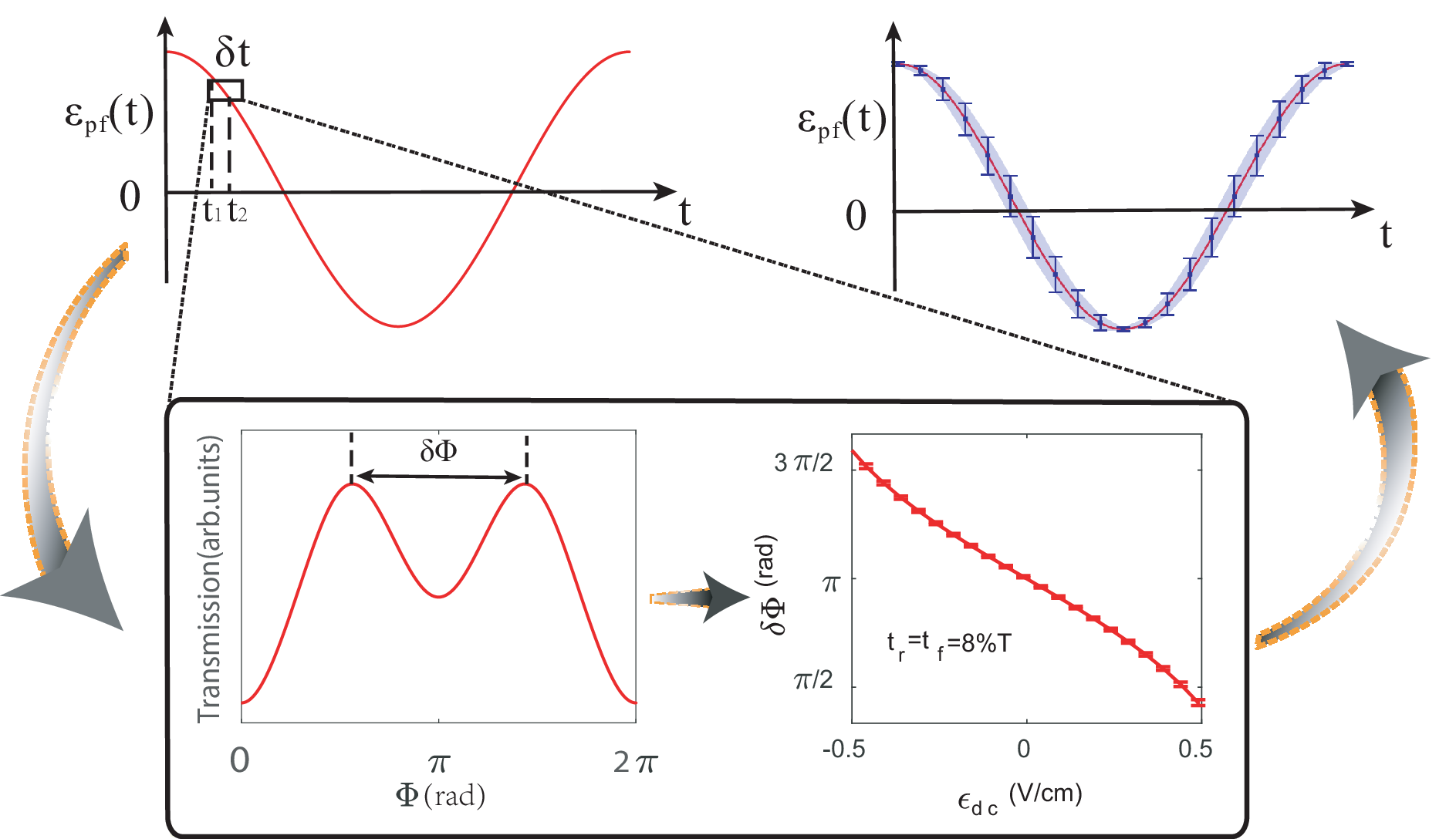}
\caption{\label{fig:4}Schematic diagram of PF field measurement. The slowly varying PF field is treated as a quasi-DC field during short-time measurements $[t_1, t_2]$. By performing repeated measurements over the time scale of the period of the PF field, complete information about the PF field can be obtained. The upper and lower bounds of the light blue shaded region correspond to the measured values obtained by treating $\epsilon_{pf}(t)|_{t=t1}$ and $\epsilon_{pf}(t)|_{t=t2}$ as the quasi-DC field in $t\in[t_1,t_2]$, respectively. The red solid line in blue shadow represents the measured values derived from using the time-averaged value of the PF field over $[t_1, t_2]$ as the quasi-DC field. Note that this is a conceptual demonstration of the measurement scheme.}
\end{figure}
Power-frequency electric fields are fundamental to the operation of electrical power systems worldwide. Accurate measurement of the PF field is crucial for various applications, including ensuring the reliability of power transmission, assessing electromagnetic interference, evaluating the impact of electric fields on biological systems and monitoring the running states of high-voltage power equipment~\cite{8444097,ZHU2019740,2024power}. Due to the sufficiently slow variation of the PF field, the DC field measurement scheme proposed in this work can be directly extended to measure PF fields.

A PF field is modeled as $\epsilon_{pf}(t) =\epsilon_{0}\text{cos}(\omega_{0}t+\Phi_{pf})$ with $\epsilon_0$, $\omega_0$ (50Hz or 60Hz) and $\Phi_{pf}$ being the amplitude, frequency and initial phase of it, respectively. The PF field period ($10^{-2}~\text{s}$) significantly exceeds the Rydberg spectrum acquisition time ($10^{-6}~\text{s}$), allowing the PF field to be treated as quasi-static during spectral measurements. In Fig.~\ref{fig:3}(a), we show the Rydberg spectra of $n=1$ bands at different times. It is shown that the $n=1$ bands indeed change dramatically with the instantaneous value of the PF field. While previous analyses assumed fixed control field detuning at the peak of the $n=1$ band, which is time-dependent, i.e., $\Delta_{c2}=-\alpha\epsilon_{rf}^2/4-\alpha[\epsilon_{pf}(t)]^2/2-\omega_g$. Here we discuss a more general situation without peak locking by fixing the control field detuning at $\epsilon_{pf}(t)$-independent point, i.e., $\Delta_{c1}=-\alpha\epsilon_{rf}^2/4-\omega_g$, as shown in Fig.~\ref{fig:3}(b). In Fig.~\ref{fig:3}(c), we compares the transmissions of the probe field as functions of the RF phase $\Phi$ for both detuning situations ($\Delta_c=\Delta_{c1}$ and $\Delta_c=\Delta_{c2}$). Notably, the peak separations $\delta\Phi$ remain identical between the two situations. In Fig.~\ref{fig:3}(d), we further demonstrate that the control field detuning ($\Delta_c=\Delta_{c1}$ or $\Delta_c=\Delta_{c2}$) does not affect the corresponding relationship between $\delta\Phi$ and $\epsilon_{dc}$ for the square-wave modulation with $t_r=t_f=0$ and $t_r=t_f=8\%\tau$.

Under suboptimal experimental conditions, we assume that obtaining the curve $T_1$ as a function of RF phase $\Phi$ as shown in Fig.~\ref{fig:3}(c) (to obtain $\delta \Phi$) requires a measurement time of $10^{-3}~\text{s}$. Within this time-frame, the PF field is treated as a quasi-DC field, and its value can be obtained by referring to the known correspondence chart between $\delta \Phi$ and $\epsilon_{dc}$ as shown in Fig.~\ref{fig:3}(d). Repeated measurements over the $10^{-2}~\text{s}$ timescale enable full reconstruction of the PF field dynamics, as shown in Fig.~\ref{fig:4}. The error bar of the PF field dynamics originate from the measurement errors of the instantaneous values of the PF field, which primarily stem from two aspects: discretizing the slowly varying PF field into a quasi-DC field and the measurement error of the DC field. The error bars are $|\epsilon_{pf}(t)|_{t=t2}-\epsilon_{pf}(t)|_{t=t1}|+\delta\epsilon_{dc}$ with $t_2-t_1=0.001$ and $\delta\epsilon_{dc}=\delta V/d$, where $\delta V$ is the accuracy for the square-wave modulation with $t_r=t_f=8\%\tau$ in Fig.~\ref{fig:2}(d) and $d=2mm$. Reasonably, the smaller the time interval $t_2-t_1$, the smaller the measurement error of PF field.

In summary, we propose a voltage measurement scheme based on the Rydberg-atom interferometry, which overcomes the drawbacks of previous measurements based on the Stark effect. Our scheme enables the measurement of weak voltages and can distinguish between positive and negative electrodes. These results can be extended to the measurement of PF field. Our method enables full reconstruction of the PF field dynamics, thereby extracting complete information of it, including frequency, amplitude, and phase. These findings provide alternative techniques for voltage and PF electric field measurements. 

See the supplementary material for the detailed derivation of the effective Hamiltonian, the Fourier coefficient of the square-wave modulation, square-wave modulation with $L=1$ and $\delta\Phi$ independent of $\Omega_c$ and $\Omega_p$.
\begin{acknowledgments}
This work is supported by the National Natural Science Foundation of China (Grant No.~12205199, No.~12375023, No.~12204428, No.~12274272, No. 12274045 and No.~12347101), National Key Research
and Development Program of China (Grants No.~2022YFA1404201), Guangdong Basic and Applied Basic Research Foundation (Grant No.~2025A1515011840), Natural Science Foundation of Top Talent of SZTU (grant No.~GDRC202312), the Natural Science Foundation of Henan Province (Grant No.~242300421159) and Guangdong Provincial Quantum Science Strategic Initiative (No.~GDZX2305006).
\end{acknowledgments}

\section*{AUTHOR DECLARATIONS}
\subsection*{Conflict of Interest}
The authors have no conflicts to disclose.	

\section*{Data Availability Statement}
The data that support the findings of this study are available from the corresponding author upon reasonable request.

\nocite{*}
%

\end{document}